\documentclass[twocolumn]{aastex61}
\usepackage{float,graphicx,amsmath,multirow,mathtools}
\usepackage{color}
\usepackage[version=3]{mhchem}
\newcommand{\mjb}{mJy~beam$^{-1}$}
\newcommand{\kms}{km~s$^{-1}$}

\begin{document}

\title{ALMA Detection of Interstellar Methoxymethanol (\ce{CH3OCH2OH})}
\author{Brett A. McGuire}
\altaffiliation{B.A.M. is a Hubble Fellow of the National Radio Astronomy Observatory}
\affiliation{National Radio Astronomy Observatory, Charlottesville, VA 22903, USA}
\affiliation{Harvard-Smithsonian Center for Astrophysics, Cambridge, MA 02138, USA}
\author{Christopher N. Shingledecker}
\affiliation{Department of Chemistry, University of Virginia, Charlottesville, VA 22903, USA}
\author{Eric R. Willis}
\affiliation{Department of Chemistry, University of Virginia, Charlottesville, VA 22903, USA}
\author{Andrew M. Burkhardt}
\affiliation{Department of Astronomy, University of Virginia, Charlottesville, VA 22903, USA}
\author{Samer El-Abd}
\affiliation{Department of Astronomy, University of Virginia, Charlottesville, VA 22903, USA}
\author{Roman A. Motiyenko}
\affiliation{Laboratoire de Physique des Lasers, Atomes, et Mol\'ecules, UMR CNRS 8523, Universit\'e de Lille I, F-59655 Villeneuve d'Ascq C\'edex, France}
\author{Crystal L. Brogan}
\affiliation{National Radio Astronomy Observatory, Charlottesville, VA 22903, USA}
\author{Todd R. Hunter}
\affiliation{National Radio Astronomy Observatory, Charlottesville, VA 22903, USA}
\author{Laurent Margul\`{e}s}
\affiliation{Laboratoire de Physique des Lasers, Atomes, et Mol\'ecules, UMR CNRS 8523, Universit\'e de Lille I, F-59655 Villeneuve d'Ascq C\'edex, France}
\author{Jean-Claude Guillemin}
\affiliation{Institut des Sciences Chimiques de Rennes, Ecole Nationale Sup\'erieure de Chimie de Rennes, CNRS, UMR 6226, 11 All\'ee de Beaulieu, CS 50837, 35708 Rennes Cedex 7, France}
\author{Robin T. Garrod}
\affiliation{Department of Chemistry, University of Virginia, Charlottesville, VA 22903, USA}
\affiliation{Department of Astronomy, University of Virginia, Charlottesville, VA 22903, USA}
\author{Eric Herbst}
\affiliation{Department of Chemistry, University of Virginia, Charlottesville, VA 22903, USA}
\affiliation{Department of Astronomy, University of Virginia, Charlottesville, VA 22903, USA}
\author{Anthony J. Remijan}
\affiliation{National Radio Astronomy Observatory, Charlottesville, VA 22903, USA}
\correspondingauthor{Brett A. McGuire}
\email{bmcguire@nrao.edu}

\begin{abstract}
\noindent We report the detection of interstellar methoxymethanol (\ce{CH3OCH2OH}) in ALMA Bands~6 and~7 toward the MM1 core in the high-mass star-forming region NGC 6334I at $\sim$0.1\arcsec~--~1\arcsec~spatial resolution.  A column density of $4(2)\times10^{18}$~cm$^{-2}$ at $T_{ex}$~=~200~K is derived toward MM1, $\sim$34~times less abundant than methanol (\ce{CH3OH}), and  significantly higher than predicted by astrochemical models.  Probable formation and destruction pathways are discussed, primarily through the reaction of the \ce{CH3OH} photodissociation products, the methoxy (\ce{CH3O}) and hydroxymethyl (\ce{CH2OH}) radicals.  Finally, we comment on the implications of these mechanisms on gas-phase vs grain-surface routes operative in the region, and the possibility of electron-induced dissociation of \ce{CH3OH} rather than photodissociation.
\end{abstract}

\section{Introduction}
\label{intro}

Given its high abundance in the interstellar medium (ISM), the photodissociation of methanol (\ce{CH3OH}) -- producing the methyl (\ce{CH3}), hydroxymethyl (\ce{CH2OH}), and methoxy radicals (\ce{CH3O}) -- is one of the most dominant sources of reactive organic species driving interstellar chemical evolution.  The branching ratio for these reactions,
\begin{eqnarray}
				\label{methyl}
				&\hspace{0.2in}\rightarrow\hspace{0.2in}& \ce{CH3} + \ce{OH},	\\
				\label{methoxymethyl}
\ce{CH3OH} + h\nu	&\hspace{0.2in}\rightarrow\hspace{0.2in}& \ce{CH2OH} + \ce{H},	\\
				\label{methoxy}
				&\hspace{0.2in}\rightarrow\hspace{0.2in}& \ce{CH3O} + \ce{H},	
\end{eqnarray}
is a topic of considerable interest, both in the laboratory and in astrochemical models \citep{Laas:2011yd}.  These ratios are difficult to experimentally constrain, however, given that \ce{CH2OH} and \ce{CH3O} are indistinguishable by mass spectroscopic techniques, and gas-phase production rates are low enough that direct detection by rotational spectroscopy is challenging \citep{Laas:2013bq}.  There is also a question of the relative importance of these branching ratios in gas-phase vs grain-surface chemistry \citep{Laas:2011yd}. The observation of interstellar species which are likely to be products of one or more of these radicals, and insight into its formation in either the gas or solid phase (or both), is highly desirable.

One such highly-desirable target is methoxymethanol (\ce{CH3OCH2OH}), which is thought to form primarily through either the direct reaction of \ce{CH3O} with \ce{CH2OH} in the solid-phase \citep{Garrod:2008tk}:
\begin{equation}
    \ce{CH3O} + \ce{CH2OH} \rightarrow \ce{CH3OCH2OH}
    \label{association}
\end{equation}or via an O($^1$D) insertion into dimethyl ether (\ce{CH3OCH3}). Although \ce{CH3O} has been detected in the ISM \cite{Cernicharo:2012eq}, \ce{CH2OH} has proven elusive, making it difficult to accurately incorporate into chemical models.  If \ce{CH3OCH2OH} does form through Reaction~\ref{association}, and the rate can be reliably measured in the laboratory, observation of the species in the ISM would therefore provide quantitative insight into the presence of \ce{CH2OH}, even without its direct detection.

Observation of \ce{CH3OCH2OH} would therefore provide much-needed constraints on the branching ratios for Reactions~\ref{methyl}, \ref{methoxymethyl}, and \ref{methoxy}, if \ce{CH3OCH2OH} can be determined to have a grain-surface formation pathway.

Here, we present the first interstellar detection of \ce{CH3OCH2OH} using ALMA data in Bands 6 and 7 toward NGC~6334I, a massive protocluster \citep{Hunter:2006th} currently forming numerous massive stars and harboring two distinct regions of hot core line emission \citep[MM1 and MM2,][]{Brogan:2016cy,Zernickel:2012hx} at a distance of 1.3 kpc \citep{Reid:2014km}.  The observations are detailed in \S\ref{observations}, and the spectroscopy in \S\ref{spectroscopy}.  In \S\ref{results} the results are presented and analyzed, and in \S\ref{discussion}, the formation and destruction chemistry for \ce{CH3OCH2OH} is explored in the context of the detection.

\section{Observations}
\label{observations}

\begin{deluxetable*}{lcccc}
\tabletypesize{\scriptsize}
\tablecaption{Summary of ALMA observing parameters\label{obs}} 
\tablehead{\colhead{Parameter}                      & \colhead{249 GHz}        & \colhead{287 GHz}        & \colhead{303 GHz}      & \colhead{344 GHz}}  
\startdata
Observation date(s)                                 & 2017 Aug 18, 19              & 2016 Jul 18, 2016 Aug 02    & 2016 Jan 17               & 2016 Jul 17, 2016 Aug 02  \\
Cycle: configuration(s)                             & 4: C40-7                    & 3: C36-4, C36-5             & 3: C36-2                     & 3: C36-4, C36-5              \\
Project code                                        & 2016.1.00383.S              & 2015.A.00022.T              & 2015.1.00150.S            & 2015.A.00022.T            \\
Phase Center (J2000 RA, Dec)                        & 17:20:53.30, $-$35:47:00.0   & 17:20:53.36, $-$35:47:00.0   & 17:20:53.35, $-$35:47:01.5   & 17:20:53.36, $-$35:47:00.0   \\
Band                                                & Band 6                      & Band 7                      & Band 7                    & Band 7                 \\
Time on Source (min)                                & 37, 37                      & 26, 26                      & 27                        & 27, 27                 \\
Number of antennas                                  & 42, 42                      & 40, 39                      & 40                        & 38, 37                    \\
FWHP primary beam ($\arcsec$)                       & 23                          & 20                          & 19                        & 17                        \\
SPW center frequencies (GHz)                          & 239.69, 241.16, 257.09, 258.29 & 280.1, 282.0, 292.1, 294.0 & 301.20, 302.00, 303.72 & 337.1, 339.0, 349.1, 351.0\\
Bandwidth per SPW (GHz)                               & $4\times 0.9375$     &   $4\times 1.875$         & 1.875, 1.875, 0.46875        & $4\times 1.875$    \\
Correlated channel width (MHz)                      & 0.244, 0.488, 0.244, 0.244  & 0.977                       & 0.488                     & 0.977                     \\
Bandpass calibrator                                 & J1924-2914                  & J1924-2914, J1517-2422      & J1924-2914                & J1924-2914             \\
Gain calibrator                                     & J1733-3722                  & J1717-3342                  & J1717-3342                & J1717-3342             \\
Flux calibrator                                     & J1733-1304, J1617-5848      & J1733-1304                  & J1733-1304                & J1733-1304                \\
Angular Resolution ($\arcsec\times\arcsec$ (P.A.$\arcdeg$)) & $0.11\times 0.08$ ($+76$) & $0.25\times 0.19$ ($-82$)   & $0.87\times 0.64$ ($+74$) & $0.22\times 0.17$ ($-76$)\\
Spectral Resolution (\kms\/)                              & 0.6                         & 1.1                         & 1.0                       & 1.1 \\
RMS per channel (\mjb\/ (K))\tablenotemark{a} & 1.1 (2.5)                   & 2.0 (0.62)                  & 20.0 (0.48)               & 3.3 (0.91)                         \\
\enddata
\tablenotetext{a}{The rms noise per channel varies significantly depending on whether bright emission is present in a particular channel, these estimated noise levels are the average of several channels, and several off-source locations in channels with moderate line emission present.}
\end{deluxetable*}

We present detections of the \ce{CH3OCH2OH} molecule in four ALMA datasets. The salient observing and imaging parameters are presented in Table~\ref{obs}. The Cycle 3 data (observed in 2016) were calibrated by the ALMA Cycle 4 pipeline (CASA 4.7.2) as described in \citet{Hunter:2017th}, and the Cycle 4 data (observed in 2017) were calibrated by the ALMA Cycle 5 pipeline (CASA 5.1.1). For each dataset, the (relatively) line-free continuum channels were carefully selected and used to construct an initial continuum image model that was then used to iteratively self-calibrate the data. The same channels were also used to subtract the continuum emission in the uv-plane before the spectral cubes were imaged. The data presented here have been corrected for primary beam attenuation.

\section{Spectroscopy}
\label{spectroscopy}

The rotational spectra of two conformers of\\
\noindent \ce{CH3OCH2OH} were recently published by \citet{Motiyenko:2017dwa} between 150 and 450~GHz.  The detection of the lower energy conformer, Conformer I, is reported here.  The dipole moment components determined in that study at the MP2/aug-cc-pVTZ level of theory and basis set were small ($\mu_a = 0.22$~D, $\mu_b = 0.08$~D, $\mu_c = 0.13$~D).  A careful analysis of the relative intensities of A- and C-type lines in the laboratory spectrum, by a subset of the authors, showed that better agreement with corresponding relative intensities in the calculated spectrum was obtained by scaling the value of $\mu_c$ to 0.11 D.  We have therefore adopted this value for this work.   The overall RMS deviation of the fit was reported to be 40~kHz, corresponding to $\sim$0.04~km~s$^{-1}$ at these frequencies.

\section{Results and Analysis}
\label{results}

\begin{figure*}[ht!]
\centering
\includegraphics[height=0.95\textheight]{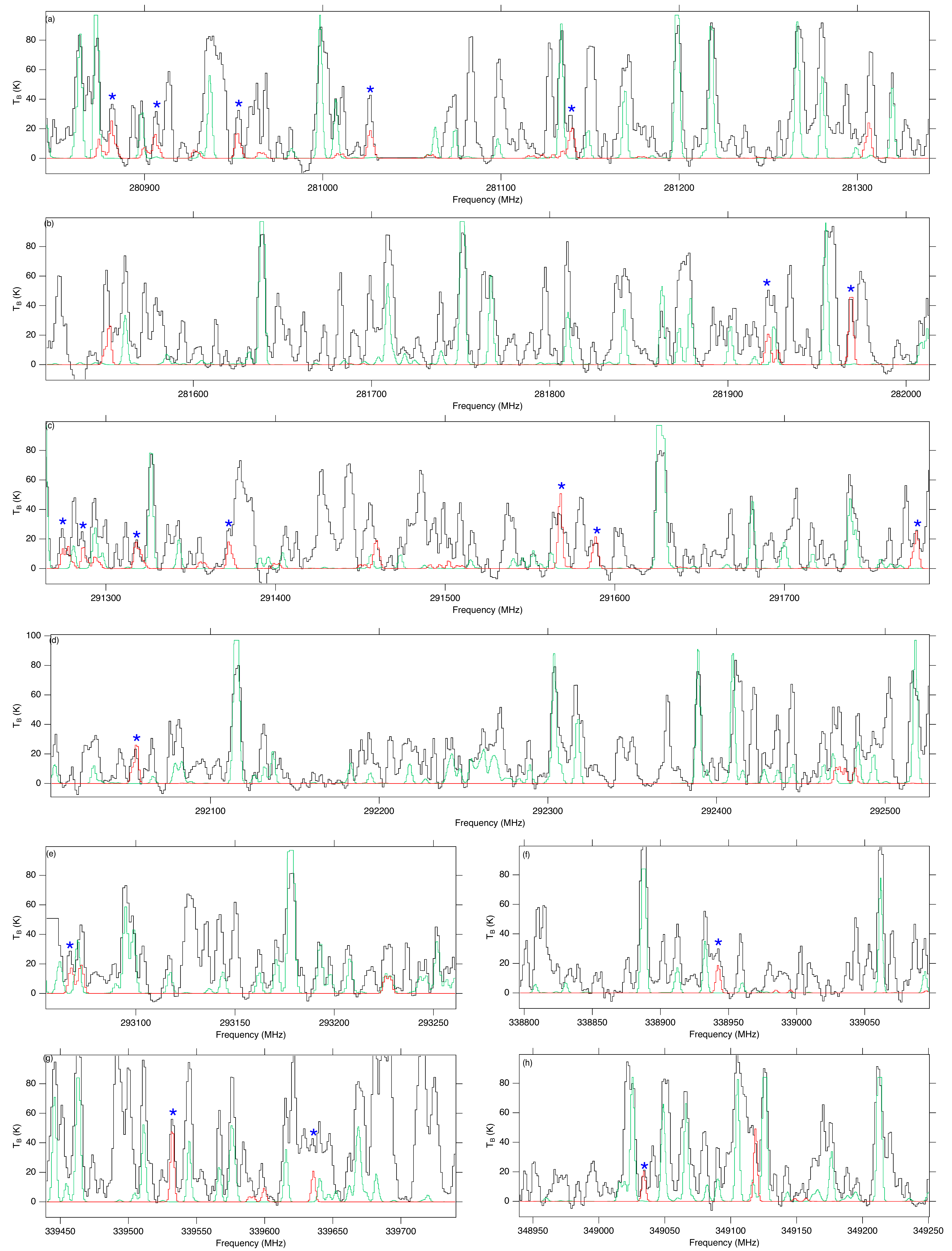}
\caption{Band~7 spectrum of NGC~6334I extracted toward the MM1 core at 0.2\arcsec~resolution in black, and adjusted to a $v_{lsr}$~=~-7~km~s$^{-1}$.  Single-excitation temperature model spectra of \ce{CH3OCH2OH} at $T_{ex}$~=~200~K, $\Delta V$~=~2.4~km~s$^{-1}$, and $N_T$~=~4(2)$\times$10$^{18}$~cm$^{-2}$ are overlaid in red.  In green are simulations of the species that are major contributors to the line density in these windows (see text \S\ref{results}).  Transitions of \ce{CH3OCH2OH} that are largely unblended (have distinguishable lineshapes) are marked with blue asterisks.}
\label{band7_hunter_detects}
\end{figure*}

Figure~\ref{band7_hunter_detects} shows the detections of \ce{CH3OCH2OH} in the 0.2\arcsec~Band 7 data (black) toward MM1 convolved to the spectral resolution of the observations, and with a 2.4~km~s$^{-1}$ linewidth (red). The position chosen for analysis (J2000 17:20:53.373, $-35$:46:58.14) lies $\sim$400~au west of the brightest continuum peak, denoted MM1b by \citet{Brogan:2016cy}.  The spectra were converted from Jy/beam to K intensity scale in each spectral window using the beam sizes listed in Table~\ref{obs}. Also shown is a total simulation in green of the major contributors to the line density in these portions of the spectrum: methyl cyanide (\ce{CH3CN})~$v_8$=0,1, \ce{CH3OH}~$v_t$=0,1, methyl formate (\ce{CH3OCHO})~$v_t$=0,1, ethanol (\ce{CH3CH2OH}), g$'$Ga-ethylene glycol (g$'$Ga-\ce{(CH2OH)2}), and g$'$Gg-\ce{(CH2OH)2}, the latter of which has only recently been detected for the first time in the ISM \citep{Jorgensen:2016cq}. Emission from many of the lines of these species is optically-thick, and a correction was applied using the formalisms of \citet{Goldsmith:1999vg} and \citet{Turner:1991um}. 

This Band~7 data contained the largest number of unblended features, and so was chosen for a preliminary excitation and column density analysis.  Given the RMS noise of the observations, the somewhat significant uncertainty introduced by the continuum subtraction due to the lack of line-free channels, and the extremely large spectral dynamic range, only features at least 10$\sigma$ above the RMS, and not substantially blended with another spectral feature, were considered significant enough to be claimed as a detected line.  

The underlying substructure of the methyl rotor-split transitions results in challenging, non-Gaussian line profiles which made it difficult to extract physically-meaningful quantitative results from a rotation diagram analysis.  Instead, the convolved spectra were simulated using a single-excitation model following the formalism established in \citet{Hollis:2004uh}:
\begin{equation}
N_T = \frac{Qe^{E_u/T_{ex}}}{\frac{8\pi^3}{3k}\nu S\mu^2}\times \frac{\frac{1}{2}\sqrt{\frac{\pi}{\ln(2)}}\frac{\Delta T_A\Delta V}{\eta_B}}{1 - \frac{e^{h\nu/kT_{ex}} -1}{e^{h\nu/kT_{bg}} -1}}
\label{column}
\end{equation}
where $N_T$ is the column density (cm$^{-2}$), $E_u$ is the upper state energy (K), $\Delta T_A \Delta V$  is integrated line intensity (K~cm~s$^{-1}$), $T_{ex}$ is the excitation temperature (K), $T_{bg}$ is the background continuum temperature ($\sim$53~K at 287~GHz and $\sim$66~K at 344~GHz), $\nu$ the transition frequency (Hz), $S$ is the intrinsic line strength, $\mu^2$ is the transition dipole moment (Debye$^2$)\footnote{These units must be properly converted to Joules$\cdot$cm$^3$ to give $N_T$ in cm$^{-2}$.}, and $\eta_B$ is the beam efficiency (assumed to be unity for these interferometric observations). The rotational partition function, $Q_{rot}$, is calculated explicitly by direct summation of states ($Q_{rot}$[300~K]~=~81794); the vibrational partition function correction, $Q_{vib}$, was obtained from quantum chemical calculations ($Q_{vib}$[300~K]~=~5.33; see Appendix~\ref{vib_part}). We assume that the source fills the beam.  The excitation temperature was fixed to 200~K, and the column density was varied to minimize the RMS intensity error for unblended transitions. A few of the strongest lines are marginally affected by optical depth, and a correction was applied using the formalisms of \citet{Goldsmith:1999vg} and \citet{Turner:1991um}.  The final derived value is $N_T$~=~4(2)$\times$10$^{18}$~cm$^{-2}$ at $T_{ex}$~=~200~K.  The $\sim$60\% uncertainty in the column density is taken as a contribution of 30\% resulting from the fit to the unblended transitions, and an estimated 50\% from the assumed excitation temperature, added in quadrature.

\begin{figure}[ht!]
\centering
\includegraphics[width=0.45\textwidth]{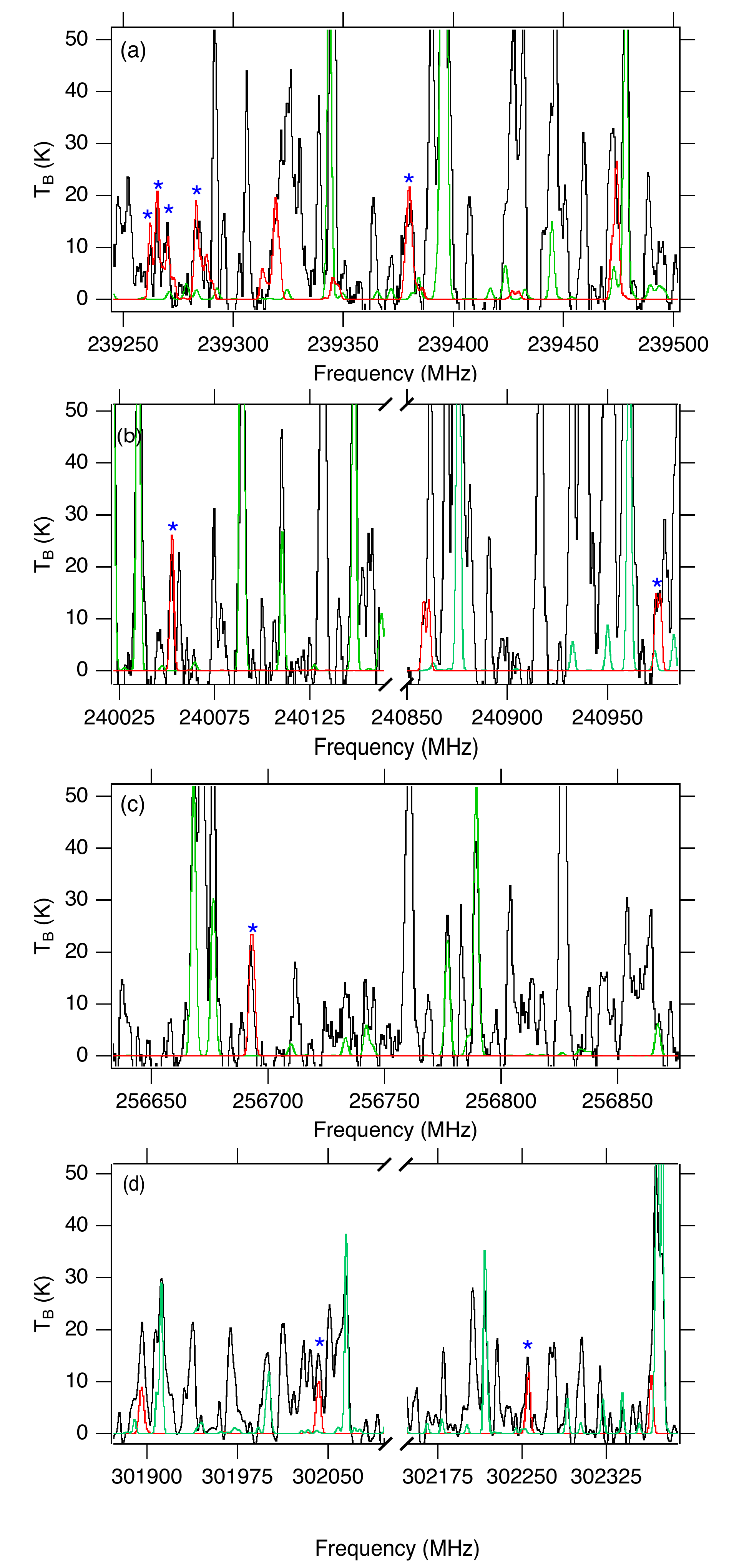}
\caption{Detections of \ce{CH3OCH2OH} indicated in red are overlaid on spectra of NGC~6334I toward MM1 in Band 6 (0.1\arcsec; panels a-c) and Band 7 (1\arcsec; panel d) in black.  The Band 6 data were adjusted to a $v_{lsr}$~=~-7~km~s$^{-1}$, while the Band 7 data were adjusted to a $v_{lsr}$~=~-8~km~s$^{-1}$ due to the different gas being probed by the larger beam.  \ce{CH3OCH2OH} simulations are at $T_{ex}$~=~200~K, $\Delta V$~=~2.4~km~s$^{-1}$ in both datasets, with $N_T$(Band~6)~=~4(2)$\times$10$^{18}$~cm$^{-2}$ and $N_T$(Band~7)~=~2$\times$10$^{18}$~cm$^{-2}$.  In green are simulations of the species that are major contributors to the line density in these windows (see text \S\ref{results}).  Transitions of \ce{CH3OCH2OH} that are largely unblended (have distinguishable lineshapes) are marked with blue asterisks.}
\label{band6_7_detects}
\end{figure}

Further unblended transitions were identified in the 0.1\arcsec~Band~6 and 1\arcsec~Band~7 datasets toward MM1 (Figure~\ref{band6_7_detects}).  The simulated profiles of \ce{CH3OCH2OH} in the Band~6 data were made using the same values of $N_T$, $T_{ex}$, and $\Delta V$ as the 0.2\arcsec~Band~7 data, and with $T_{bg}$ values appropriate for the data ($\sim$32~K).  Because these profiles show remarkably good agreement even at twice the spatial resolution, this is likely indicating that the two observations are probing the same gas, and that the distribution is at least mostly resolved.  The 1\arcsec~Band~7 data required a column density a factor of two lower, and substantially more optically thin, in agreement with a larger beam which did not resolve the substructure seen in the higher-resolution observations.  In total, more than two dozen largely unblended transitions were identified in the datasets for which a distinguishable lineshape contribution was seen in the spectrum matching the predicted shape at the correct frequency.  As noted earlier, the intensity match is not always precise, due to the uncertainty in the excitation temperature.

\subsection{Spatial Distributions}

\begin{figure*}[ht!]   
\includegraphics[width=0.99\linewidth]{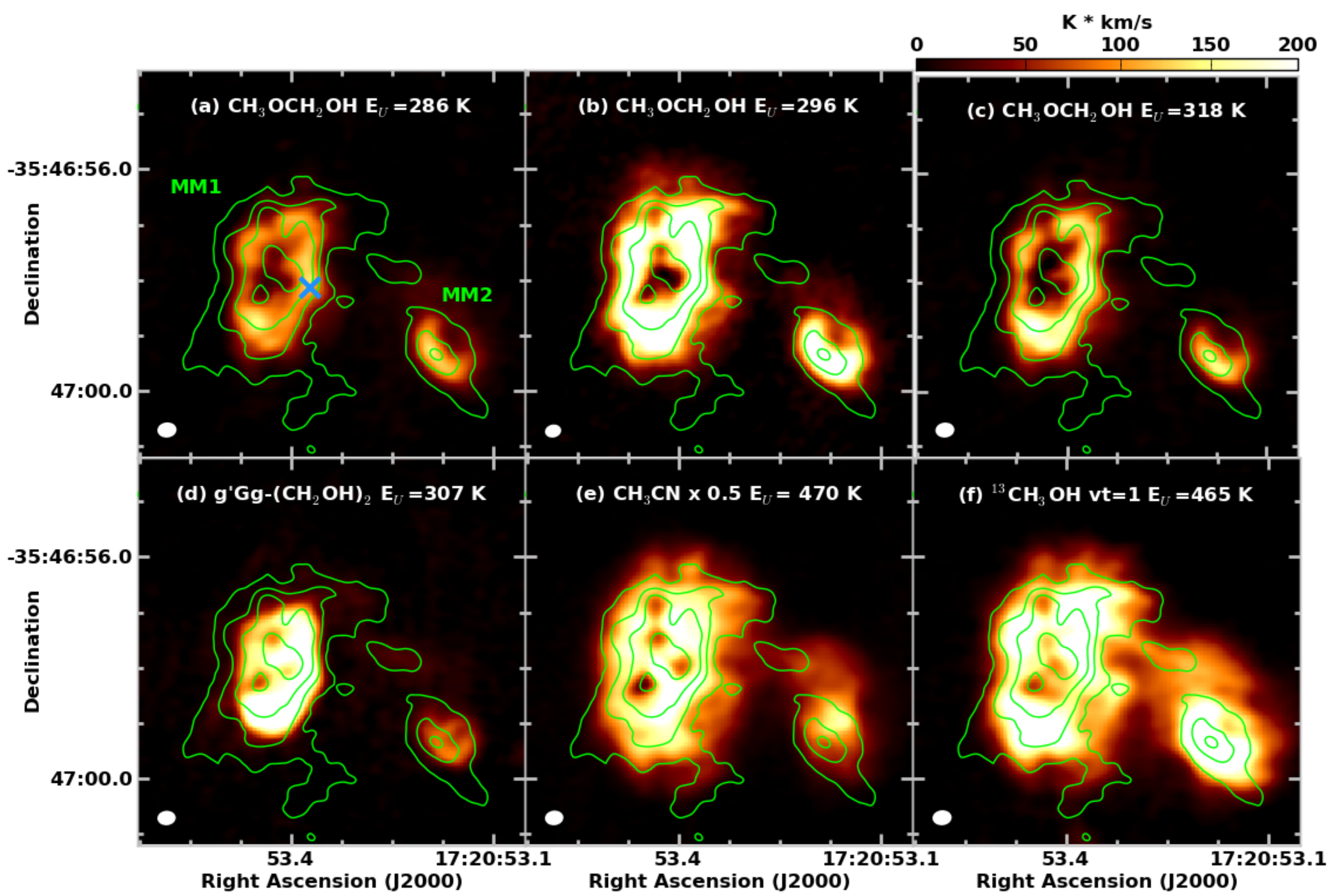}
\caption{Moment 0 maps of three \ce{CH3OCH2OH} transitions (panels $a$-$c$) compared to those of a$'$Gg-\ce{(CH2OH)2}, \ce{CH3CN}, and \ce{^{13}CH3OH}~$v_t$=1, with the 1~mm continuum overlaid as green contours (levels are 18, 53, 140, and 370 mJy beam$^{-1}$). The pertinent transition parameters for these lines are given in Table~\ref{lines} and the Supplementary Information.  The blue cross in panel (a) indicates the position from which spectra were extracted toward MM1, and the synthesized beams are shown in the lower left of each panel in white. 
\label{moments}}
\end{figure*}

The line density in the region is such that it is challenging to select transitions which are not blended from which spatial distributions can be inferred.  Figure~\ref{moments} shows integrated intensity images of three unblended \ce{CH3OCH2OH} transitions (panels a-c), as well as images of a single unblended transition each of a$'$Gg-\ce{(CH2OH)2}, \ce{CH3CN}, and \ce{^{13}CH3OH}~$v_t$=1.  The \ce{CH3OCH2OH} transitions show nearly identical spatial distributions to each other, as expected, and markedly different distributions from g$'$Gg-\ce{(CH2OH)2}, which is significantly more compact, as well as \ce{CH3CN} which, especially in MM2, shows an anti-correlated distribution.  Notably, the distribution of \ce{CH3OCH2OH} appears to be similar to, but slightly more compact than, that of \ce{^{13}CH3OH}~$v_t$=1. The ring-like appearance of the molecular emission toward MM1 is due to the high dust continuum opacity ($> 1$) and brightness temperature of the 1~mm dust continuum emission toward the dust peaks \citep[$> 100$~K][]{Brogan:2016cy}. Toward the continuum peaks, the high dust opacity attenuates the line emission from the backside of the region, and the high background continuum brightness temperature compared to the line excitation temperature of the transitions shown here (of order 100-200~K) leads to weaker line emission (lower excitation lines are seen in absorption against the continuum peaks).  To mitigate these effects, the spectra were extracted and analyzed at a position offset from the continuum peak.

\subsection{Chemical Modeling}
\label{sec:models}

\ce{CH3OCH2OH} was included in the three-phase chemical kinetics model {\it MAGICKAL} of \citet{Garrod:2013id}.  In an initial effort to explore the ability of the model to reproduce the observed abundance, we modified the model with the back-diffusion correction of \citet{Willis2017}, and used a network based on that of \citet{Belloche:2017}, in which \ce{CH3OCH2OH} is produced via Reaction~\ref{association}. The network also contains likely gas-phase and grain-surface destruction mechanisms for \ce{CH3OCH2OH}. 

The chemical modeling was performed using the two-stage approach described in \citet{Garrod:2013id}. Phase 1 is a cold collapse to a maximum density of 2 x 10$^8$ cm$^{-3}$ and a dust grain temperature of 8 K. This is followed by a warm-up from 8 K to 400 K, simulating the `ignition' of a hot core. Three different warm-up timescales have been used, as the choice of timescale has been previously shown to have significant effects on the chemistry \citep{Garrod:2013id}. The fastest timescale reaches a temperature of 200 K at 5 x 10$^4$ yr, and 400 K at 7.12 x 10$^4$ yr. The slowest timescale reaches these milestones at 10$^6$ yr and 1.43 x 10$^6$ yr; the intermediate timescale takes 2 x 10$^5$ yr and 2.85 x 10$^5$ yr.  

\begin{figure*}[ht!]
\centering
\includegraphics[width=0.99\textwidth]{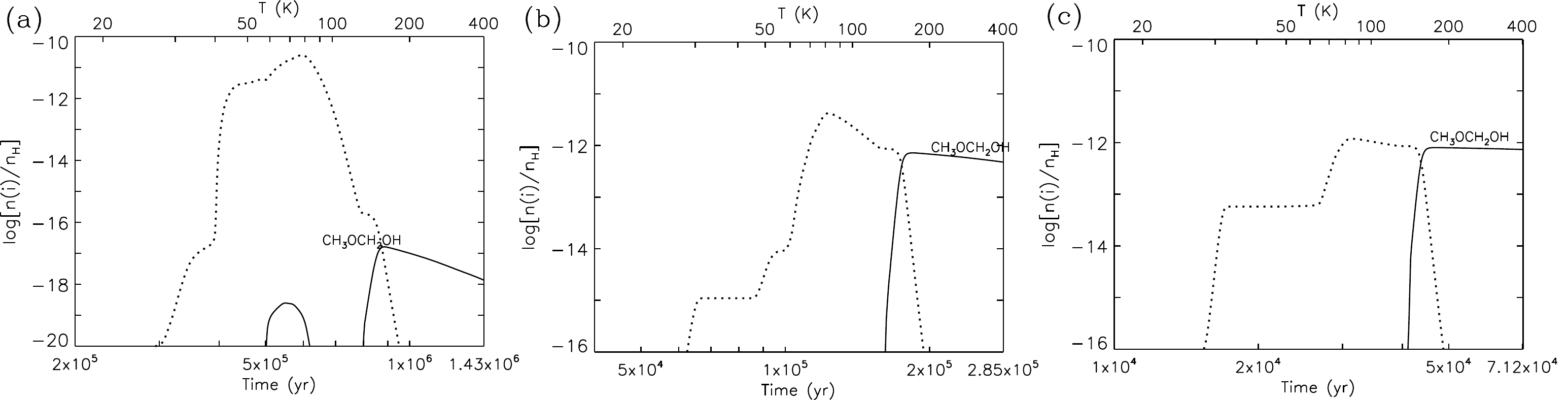}
\caption{Abundance profiles of \ce{CH3OCH2OH} for three warm-up timescales. Gas-phase abundance is displayed as a solid curve, while grain-surface abundance is shown as a dotted curve. Panel (a) shows the abundance profile for the longest warm-up timescale; (b) and (c) show the abundance profile for the intermediate and fast warm-up timescales, respectively.}
\label{model}
\end{figure*}

The results of the model are shown graphically in Figure~\ref{model}. The fast warm-up timescale produces the highest abundance of \ce{CH3OCH2OH}, with a maximum fractional abundance of $\sim$10$^{-12}$ with respect to total hydrogen, which is well-maintained to a temperature of 400 K. The abundance of \ce{CH3OH} in the same model approaches 10$^{-5}$, corresponding to \ce{CH3OCH2OH}:\ce{CH3OH} of $\sim$10$^{-7}$.  The observed ratio of \ce{CH3OCH2OH}:\ce{CH3OH} is estimated to be 1:34 (see Appendix~\ref{ch3oh_column}). The intermediate timescale shows very similar results, while the slowest timescale produces a very low gas-phase abundance of \ce{CH3OCH2OH}, peaking at $\sim$10$^{-17}$. This is due to the increased efficiency of grain-surface destruction mechanisms in the slow warm-up timescale.
\vspace{3em}

\section{Discussion}
\label{discussion}

Most complex organic molecules (COMs), especially saturated species, are thought to have efficient formation pathways in the solid-phase on the surface of dust grains \citep{Garrod:2008tk}.  In star-forming regions like NGC~6334I, the gradual warm-up of these grains can then thermally desorb the product molecules into the gas phase, where they are detected by radio observations \citep{Garrod:2013id}.  For instance, the successive hydrogenation of CO on grains is likely the major formation route for methanol \citep{watanabe_methanol_2002,fuchs_methanol_2009}. Yet, other studies have shown that some saturated COMs, such as methyl formate and formamide (\ce{NH2CHO}), might have significant, if perhaps not dominant, gas-phase formation pathways \citep{Neill:2012fr,Codella:2017ij}.  It is worthwhile, then, to consider the potential operative pathways, both in the gas- and solid-phase, for the formation of \ce{CH3OCH2OH}.


\subsection{Radical-Radical Recombination}

As noted by \citet{Motiyenko:2017dwa}, the most obvious astrochemically relevant formation mechanism is the association of the \ce{CH3O} and \ce{CH2OH} radicals, which can occur on grains or in the gas.  In the gas phase, an association reaction occurs via an intermediate complex, which re-dissociates into reactants unless it can relax sufficiently rapidly.  In the low density interstellar medium, the relaxation mechanism is radiative, typically via emission from vibrational levels above the energy of reactants to levels below this energy and therefore stable.  For ion-neutral systems such as \begin{equation}
    \ce{CH3+} + \ce{H2} \rightarrow \ce{CH5+} + h\nu,
    \label{ch5plus}
\end{equation}experiments in ion traps and other low-density devices have been performed and show a wide range of reaction rate coefficients, depending upon the depth of the complex potential well (which helps to determine the lifetime of the complex against dissociation), the size of the reactants, and the temperature \citep{Gerlich:1992du}.  The experiments have also been supplemented by statistical calculations (e.g. \citealt{Herbst:1985eh}).   

Both approaches show that a large complex well depth (a few eV), large reactants (more than 4-5 atoms), and low temperatures ($<$100 K) are needed for the radiative rate coefficient to approach the collisional rate coefficient, and make these reactions feasible.  Without these constraints, much smaller rate coefficients are obtained.  

For neutral-neutral systems, very few reactions have been studied in the laboratory.  Some theoretical rate coefficients have been obtained via statistical theories \citep{Vuitton:2011ks}; these indicate indirectly that the association of the \ce{CH3O} and \ce{CH2OH} radicals might have a large rate coefficient at temperatures under 100 K, although it is also likely that the low gas-phase abundances of the reactants would render the process inefficient even with a large rate coefficient.  Thus, if the radicals \ce{CH3O} and \ce{CH2OH} are more abundant on grains than in the gas for large ranges of temperature, the solid-phase association should be the more important. On the other hand, in their gas-grain modeling of the formation of \ce{CH3OCH3} in cold clouds, \citet{Balucani:2015gj} found the dominant process to be the gas-phase radiative association between \ce{CH3} and \ce{CH3O}, assuming that it occurred at the collisional rate.

\subsection{O($^1$D) Insertion Reactions}

Another possible formation route involves the insertion of O($^1$D) into one of the C-H bonds of \ce{CH3OCH3} \citep{Hays:2013fn}. Electronically excited species like O($^1$D) can form in cosmic ray-irradiated dust-grain ice-mantles, where they quickly react with a neighboring species or are quenched by the solid \citep{shingledecker2017a,shingledecker2017b}.  Indeed, \citet{bergner_methanol_2017} recently found evidence that an analogous reaction, \begin{equation}
\mathrm{O(^1D)} + \mathrm{CH_4} \rightarrow \mathrm{CH_3OH,} 
\label{osinglet}
\end{equation} could occur efficiently in interstellar ices.  It is possible that such excited species could also react in the gas-phase in the ISM, however, it is perhaps more likely that the excited species will emit a photon and radiatively relax before encountering a collisional reaction partner in the gas phase.

\subsection{Photodissociation vs Cosmic Rays}

As mentioned in \S1, both \ce{CH3O} and \ce{CH2OH} can be formed from the dissociation of methanol. In the ISM, the two main drivers of dissociation are photons and cosmic rays. As recently noted by \citet{shingledecker2017b}, the products of photochemistry and radiation chemistry are often similar, but not necessarily identical, due to differences in the underlying microscopic interactions that drive such processes. The distinction between photochemistry and radiation chemistry is potentially of importance for \ce{CH3OCH2OH}, and was noted by \citet{boamah_low-energy_2014} and \citet{sullivan_low-energy_2016}. In the experiments described in those works, \ce{CH3OCH2OH} was detected after the exposure of condensed methanol to low-energy ($<$20 eV) electrons, which are characteristic of the secondary electrons produced in interstellar ices due to cosmic ray bombardment \citep{shingledecker2017a}. In photodissociation experiments, however, \ce{CH3OCH2OH} was not detected as a product of UV-irradiated methanol ice, which \citet{sullivan_low-energy_2016} indicate might make \ce{CH3OCH2OH} a good tracer of cosmic ray-induced chemistry in the ISM.

\subsection{Grain-Surface Hydrogenation}

Yet another potential formation pathway is through hydrogenation of precursor species on grain surfaces, such as:\begin{equation}
\ce{CH3O} + \ce{CH2O} \rightarrow \ce{CH3OCH2O} \xrightarrow[]{\ce{H}} \ce{CH3OCH2OH}
\end{equation} or
\begin{equation}
    \ce{CH3OCHO} \xrightarrow[]{\ce{H}} \xrightarrow[]{\ce{H}} \ce{CH3OCH2OH.}
\end{equation}
Successive hydrogenation reactions have been proposed as the formation mechanism for a number of saturated species, including \ce{CH3OH} \citep{Woon:2002wu} and  \ce{NH2OH} \citep{Fedoseev:2016gt} among numerous others \cite{Linnartz:2015ec}, although recent work has suggested this may not be efficient for substituted aldehydes \citep{Jonusas:2017ga}.  A more in-depth analysis of the potential of these types of reactions to both form \ce{CH3OCH2OH} and suggest other potential products, is warranted.

\subsection{Comparison to Models}

As discussed in \S\ref{sec:models}, current models using only Reaction~\ref{association} are significantly under-producing \ce{CH3OCH2OH}.  While the derived abundance of \ce{CH3OH} in our observations has a high degree of uncertainty, it is clear that the current models do not properly treat \ce{CH3OCH2OH} formation, destruction, or both.  Yet, as shown in Figure~\ref{moments}, the \ce{CH3OCH2OH} distribution is remarkably similar to that of \ce{CH3OH}, and different from other COMs.  While not definitive, this is highly suggestive of a closely-tied chemistry between these two species that would favor Reaction~\ref{association}.

\section{Conclusions}
\label{conclusions}

We have reported the detection of methoxymethanol \ce{CH3OCH2OH} for the first time in the ISM using ALMA Bands 6 and 7 observations toward the massive Galactic protocluster NGC~6334I.  More than two dozen unblended transitions above 10$\sigma$ were identified.  These transitions show identical spatial distributions, distinct from other complex species, and well-matched to that of \ce{CH3OH}, a likely precursor.  Current treatments for the grain-surface formation of \ce{CH3OCH2OH} from \ce{CH3O} and \ce{CH2OH} significantly under-produce \ce{CH3OCH2OH}, likely indicating other unconsidered formation pathways exist.  Further, as has recently been suggested, cosmic ray-induced chemistry may play a substantial role, making \ce{CH3OCH2OH} a potentially powerful tracer of this process.

\acknowledgments

This paper makes use of the following ALMA data: ADS/JAO.ALMA\#2015.A.00022.T,  \\ ADS/JAO.ALMA\#2015.1.00150.S, and \\ ADS/JAO.ALMA\#2016.1.00383.S. ALMA is a partnership of ESO (representing its member states), NSF (USA) and NINS (Japan), together with NRC (Canada) and NSC and ASIAA (Taiwan) and KASI (Republic of Korea), in cooperation with the Republic of Chile. The Joint ALMA Observatory is operated by ESO, AUI/NRAO and NAOJ.   The National Radio Astronomy Observatory is a facility of the National Science Foundation operated under cooperative agreement by Associated Universities, Inc.  E. H. thanks the National Science Foundation for support of his astrochemistry program.  A.M.B. is a Grote Reber Fellow, and support for this work was provided by the NSF through the Grote Reber Fellowship Program administered by Associated Universities, Inc./National Radio Astronomy Observatory. Support for B.A.M. was provided by NASA through Hubble Fellowship grant \#HST-HF2-51396 awarded by the Space Telescope Science Institute, which is operated by the Association of Universities for Research in Astronomy, Inc., for NASA, under contract NAS5-26555. This work was supported by the Programme National ``Physique et Chimie du Milieu Interstellaire" (PCMI) of CNRS/INSU with INC/INP co-funded by CEA and CNES.


\begin{thebibliography}{}
\expandafter\ifx\csname natexlab\endcsname\relax\def\natexlab#1{#1}\fi
\providecommand{\url}[1]{\href{#1}{#1}}

\bibitem[{Balucani {et~al.}(2015)Balucani, Ceccarelli, \&
  Taquet}]{Balucani:2015gj}
Balucani, N., Ceccarelli, C., \& Taquet, V. 2015, MNRAS, 449, L16

\bibitem[{{Belloche} {et~al.}(2017){Belloche}, {Meshcheryakov}, {Garrod},
  {Ilyushin}, {Alekseev}, {Motiyenko}, {Margul{\`e}s}, {M{\"u}ller}, \&
  {Menten}}]{Belloche:2017}
{Belloche}, A., {Meshcheryakov}, A.~A., {Garrod}, R.~T., {et~al.} 2017, \aap,
  601, A49

\bibitem[{Bergner {et~al.}(2017)Bergner, \"{O}berg, \&
  Rajappan}]{bergner_methanol_2017}
Bergner, J.~B., \"{O}berg, K.~I., \& Rajappan, M. 2017, ApJ, 845, 29.

\bibitem[{Boamah {et~al.}(2014)Boamah, Sullivan, Shulenberger, Soe, Jacob,
  Yhee, Atkinson, Boyer, Haines, \& Arumainayagam}]{boamah_low-energy_2014}
Boamah, M.~D., Sullivan, K.~K., Shulenberger, K.~E., {et~al.} 2014, Faraday
  Discussions, 168, 249.

\bibitem[{Brogan {et~al.}(2016)Brogan, Hunter, Cyganowski, Chandler, Friesen,
  \& Indebetouw}]{Brogan:2016cy}
Brogan, C.~L., Hunter, T.~R., Cyganowski, C.~J., {et~al.} 2016, ApJ, 832, 1

\bibitem[{Cernicharo {et~al.}(2012)Cernicharo, Marcelino, Roueff, Gerin,
  Jim{\'e}nez-Escobar, \& Mu{\~n}oz~Caro}]{Cernicharo:2012eq}
Cernicharo, J., Marcelino, N., Roueff, E., {et~al.} 2012, ApJ, 759, L43

\bibitem[{Codella {et~al.}(2017)Codella, Ceccarelli, Caselli, Balucani, Barone,
  Fontani, Lefloch, Podio, Viti, Feng, Bachiller, Bianchi, Dulieu,
  Jimenez-Serra, Holdship, Neri, Pineda, Pon, Sims, Spezzano, Vasyunin, Alves,
  Bizzocchi, Bottinelli, Caux, Chacon-Tanarro, Choudhury, Coutens, Favre,
  Hily-Blant, Kahane, Jaber Al-Edhari, Laas, Lopez-Sepulcre, Ospina, Oya,
  Punanova, Puzzarini, Quenard, Rimola, Sakai, Skouteris, Taquet, Testi,
  Theule, Ugliengo, Vastel, Vazart, Wiesenfeld, \& YAMAMOTO}]{Codella:2017ij}
Codella, C., Ceccarelli, C., Caselli, P., {et~al.} 2017, A\&A, 605, L3

\bibitem[{Fedoseev {et~al.}(2016)Fedoseev, Chuang, van Dishoeck, Ioppolo, \&
  Linnartz}]{Fedoseev:2016gt}
Fedoseev, G., Chuang, K.~J., van Dishoeck, E.~F., Ioppolo, S., \& Linnartz, H.
  2016, MNRAS, 460, 4297
  
\bibitem[{Frisch {et~al.}(2009)Frisch, M J and Trucks, G W and Schlegel, H B and Scuseria, G E and Robb, M A and Cheeseman, J R and Scalmani, G and Barone, V and Mennucci, B and Petersson, G A and Nakatsuji, H and Caricato, M and Li, X and Hratchian, H P and Izmaylov, A F and Bloino, J and Zheng, G and Sonnenberg, J L and Hada, M and Ehara, M and Toyota, K and Fukuda, R and Hasegawa, J and Ishida, M and Nakajima, T and Honda, Y and Kitao, O and Nakai, H and Vreven, T and Montgomery, J A and Peralta, J E and Ogliaro, F and Bearpark, M and Heyd, J J and Brothers, E and Kudin, K N and Staroverov, V N and Kobayashi, R and Normand, J and Raghavachari, K and Rendell, A and Burant, J C and Iyengar, S S and Tomasi, J and Cossi, M and Rega, N and Millam, J M and Klene, M and Knox, J E and Cross, J B and Bakken, V and Adamo, C and Jaramillo, J and Gomperts, R and Stratmann, R E and Yazyev, O and Austin, A J and Cammi, R and Pomelli, C and Ochterski, J W and Martin, R L and Morokuma, K and Zakrzewski, V G and Voth, G A and Salvador, P and Dannenberg, J J and Dapprich, S and Daniels, A D and {Farkas} and Foresman, J B and Ortiz, J V and Cioslowski, J and Fox, D J}]{gaussian09B1}Frisch, M., Trucks, G., Schlegel, H., {et~al.} 2009, Gaussian 09, Revision B.01, Gaussian, Inc., Wallingford CT

\bibitem[{{Fuchs} {et~al.}(2009){Fuchs}, {Cuppen}, {Ioppolo}, {Romanzin},
  {Bisschop}, {Andersson}, {van Dishoeck}, \& {Linnartz}}]{fuchs_methanol_2009}
{Fuchs}, G.~W., {Cuppen}, H.~M., {Ioppolo}, S., {et~al.} 2009, \aap, 505, 629

\bibitem[{Garrod(2013)}]{Garrod:2013id}
Garrod, R.~T. 2013, ApJ, 765, 60

\bibitem[{Garrod {et~al.}(2008)Garrod, Weaver, \& Herbst}]{Garrod:2008tk}
Garrod, R.~T., Weaver, S. L.~W., \& Herbst, E. 2008, ApJ, 682, 283

\bibitem[{Gerlich \& Horning(1992)}]{Gerlich:1992du}
Gerlich, D., \& Horning, S. 1992, Chem Rev, 92, 1509

\bibitem[{Goldsmith \& Langer(1999)}]{Goldsmith:1999vg}Goldsmith, P.F. \& Langer, W.D. 1999, ApJ, 517, 209.

\bibitem[{Hays \& Widicus~Weaver(2013)}]{Hays:2013fn}
Hays, B.~M., \& Widicus~Weaver, S.~L. 2013, JPCA, 117, 7142

\bibitem[{Herbst(1985)}]{Herbst:1985eh}
Herbst, E. 1985, ApJ, 291, 226

\bibitem[{Hollis {et~al.}(2004)Hollis, Jewell, Lovas, \&
  Remijan}]{Hollis:2004uh}
Hollis, J.~M., Jewell, P.~R., Lovas, F.~J., \& Remijan, A. 2004, ApJ, 613, L45

\bibitem[{Hunter {et~al.}(2006)Hunter, Brogan, Megeath, Menten, Beuther, \&
  Thorwirth}]{Hunter:2006th}
Hunter, T.~R., Brogan, C.~L., Megeath, S.~T., {et~al.} 2006, ApJ, 649, 888

\bibitem[{Hunter {et~al.}(2017)Hunter, Brogan, MacLeod, Cyganowski, Chandler,
  Chibueze, Friesen, Indebetouw, Thesner, \& Young}]{Hunter:2017th}
Hunter, T.~R., Brogan, C.~L., MacLeod, G., {et~al.} 2017, ApJL, 837, L29
  
\bibitem[{Jonusas {et~al.}(2017)Jonusas, Mindaugas, Guillemin, Jean-Claude, \& Krim, Lahouari}]{Jonusas:2017ga}Jonusas, M., Guillemin, J.-C., \& Krim, L. 2017, MNRAS, 468, 4592.

\bibitem[{J{\o}rgensen {et~al.}(2016)J{\o}rgensen, van~der Wiel, Coutens,
  Lykke, Muller, van Dishoeck, Calcutt, Bjerkeli, Bourke, Drozdovskaya,
  Fayolle, Favre, Garrod, Jacobsen, {\"O}berg, Persson, \&
  Wampfler}]{Jorgensen:2016cq}
J{\o}rgensen, J.~K., van~der Wiel, M., Coutens, A., {et~al.} 2016, A\&A, 595, A117

\bibitem[{Laas {et~al.}(2011)Laas, Garrod, Herbst, \&
  Widicus~Weaver}]{Laas:2011yd}
Laas, J.~C., Garrod, R.~T., Herbst, E., \& Widicus~Weaver, S.~L. 2011, ApJ, 728, 71

\bibitem[{Laas {et~al.}(2013)Laas, Hays, \& Widicus~Weaver}]{Laas:2013bq}
Laas, J.~C., Hays, B.~M., \& Widicus~Weaver, S.~L. 2013, JPCA, 117, 9548

\bibitem[{Linnartz {et~al.}(2015)Linnartz, Ioppolo, \&
  Fedoseev}]{Linnartz:2015ec}
Linnartz, H., Ioppolo, S., \& Fedoseev, G. 2015, International Reviews in
  Physical Chemistry, 34, 205

\bibitem[{Milam {et~al.}(2005)Milam, Savage, Brewster, Ziurys, \&
  Wyckoff}]{Milam:2005gt}
Milam, S.~N., Savage, C., Brewster, M.~A., Ziurys, L.~M., \& Wyckoff, S. 2005,
  ApJ, 634, 1126

\bibitem[{Motiyenko {et~al.}(2017)Motiyenko, Margules, Despois, \&
  Guillemin}]{Motiyenko:2017dwa}
Motiyenko, R., Margules, L., Despois, D., \& Guillemin, J.-C. 2017, PCCP, in press, doi:10.1039/c7cp05932a.

\bibitem[{Neill {et~al.}(2012)Neill, Muckle, Zaleski, Steber, Pate, Lattanzi,
  Spezzano, McCarthy, \& Remijan}]{Neill:2012fr}
Neill, J.~L., Muckle, M.~T., Zaleski, D.~P., {et~al.} 2012, ApJ, 755, 153

\bibitem[{Reid {et~al.}(2014)Reid, Menten, Brunthaler, Zheng, Dame, Xu, Wu,
  Zhang, Sanna, Sato, Hachisuka, Choi, Immer, Moscadelli, Rygl, \&
  Bartkiewicz}]{Reid:2014km}
Reid, M.~J., Menten, K.~M., Brunthaler, A., {et~al.} 2014, ApJ, 783, 130

\bibitem[{Shingledecker \& Herbst(2017)}]{shingledecker2017b}
Shingledecker, C.~N., \& Herbst, E. 2017, Phys. Chem. Chem. Phys., in press,
  doi:10.1039/C7CP05901A.

\bibitem[{Shingledecker {et~al.}(2017)Shingledecker, Le~Gal, \&
  Herbst}]{shingledecker2017a}
Shingledecker, C.~N., Le~Gal, R., \& Herbst, E. 2017, Phys. Chem. Chem. Phys.,
  19, 11043.

\bibitem[{Sullivan {et~al.}(2016)Sullivan, Boamah, Shulenberger, Chapman,
  Atkinson, Boyer, \& Arumainayagam}]{sullivan_low-energy_2016}
Sullivan, K.~K., Boamah, M.~D., Shulenberger, K.~E., {et~al.} 2016, MNRAS, 460, 664.

\bibitem[{Turner (1991)Turner, B.}]{Turner:1991um}Turner, B.E. 1991, ApJS, 76, 617.

\bibitem[{Vuitton {et~al.}(2011)Vuitton, Yelle, Lavvas, \&
  Klippenstein}]{Vuitton:2011ks}
Vuitton, V., Yelle, R.~V., Lavvas, P., \& Klippenstein, S.~J. 2011, ApJ, 744, 11

\bibitem[{{Watanabe} \& {Kouchi}(2002)}]{watanabe_methanol_2002}
{Watanabe}, N., \& {Kouchi}, A. 2002, \apjl, 571, L173

\bibitem[{{Willis} \& {Garrod}(2017)}]{Willis2017}
{Willis}, E.~R., \& {Garrod}, R.~T. 2017, \apj, 840, 61

\bibitem[{Woon(2002)}]{Woon:2002wu}
Woon, D.~E. 2002, ApJ, 569, 541

\bibitem[{Zernickel {et~al.}(2012)Zernickel, Schilke, Schmiedeke, Lis, Brogan,
  Ceccarelli, Comito, Emprechtinger, Hunter, \& M{\"o}ller}]{Zernickel:2012hx}
Zernickel, A., Schilke, P., Schmiedeke, A., {et~al.} 2012, A\&A, 546, A87

\end{thebibliography}

\appendix

\renewcommand{\thefigure}{A\arabic{figure}}
\renewcommand{\thetable}{A\arabic{table}}
\renewcommand{\theequation}{A\arabic{equation}}
\setcounter{figure}{0}
\setcounter{table}{0}
\setcounter{equation}{0}

\section{Observed Line Parameters}
\label{lineparams}

\begin{deluxetable*}{cccc}[h!]
\tabletypesize{\footnotesize}
\tablecaption{Pertinent line parameters of selected observed \ce{CH3OCH2OH}, g$'$Gg-\ce{(CH2OH)2}, \ce{CH3CN}, and \ce{^{13}CH3OH}~$v_t$=1 transitions.}
\tablehead{
\colhead{\multirow{2}{*}{Quantum Numbers}}&   \colhead{Frequency}       &   \colhead{$S_{ij}\mu^2$}	&	\colhead{$E_{u}$}     \\
                                &   \colhead{(MHz)}           &   \colhead{(Debye$^2$)}     &   \colhead{(K)}         
}
\startdata
\multicolumn{4}{c}{\ce{CH3OCH2OH}}                                                      \\
$27_{12,15} - 26_{12,14}$~E      & 	 281024.6590 	 & 	 1.05 	 & 	 271.81 	 \\ 
$27_{12,16} - 26_{12,15}$~A 	 & 	 281026.0480 	 & 	 1.05 	 & 	 271.81 	 \\ 
$27_{12,15} - 26_{12,14}$~A 	 & 	 281026.0480 	 & 	 1.05 	 & 	 271.81 	 \\ 
$27_{12,16} - 26_{12,15}$~E 	 & 	 281027.8190 	 & 	 1.05 	 & 	 271.81 	 \\ 
\multicolumn{4}{c}{g$'$Gg-\ce{(CH2OH)2}\tablenotemark{$^{\dagger}$} }                                                     \\
$28_{15,13} - 28_{14,14}$~$v_t$=1&  292703.3917     &   199.61  &   307.44      \\
$28_{15,14} - 28_{14,15}$~$v_t$=1&  292703.3917     &   155.24  &   307.44      \\
\multicolumn{4}{c}{\ce{CH3CN}}                                                     \\
$16_{7} - 15_{7}$               &   294.02549      &   274.21  &   469.79  \\
\multicolumn{4}{c}{\ce{^{13}CH3OH}~$v_t$=1 }                                                     \\
$6_{3,4}  -5_{3,3}$             &   282383.0330     &   3.64     &  465 \\
\enddata
\tablenotetext{\dagger}{Here, $v_t$ refers to a large-amplitude tunneling motion of  the OH moieties.}
\tablecomments{Only a sample of the \ce{CH3OCH2OH} transitions is shown here; the full table of all transitions within the range of these observations is available as Supplementary Information.}
\label{lines}
\end{deluxetable*}

\section{\ce{CH3OCH2OH} Vibrational Partition Function}
\label{vib_part}

The vibrational correction to the partition function for \ce{CH3OCH2OH} was calculated according to Equation~\ref{vib_part_eq}:
\begin{equation}
    Q(T)_{vib}=\prod\limits_{\substack{i=1}}^{3N-6} \frac{1}{1-e^{-E_i/kT}}.
    \label{vib_part_eq}
\end{equation}
To obtain the vibrational energy levels for \ce{CH3OCH2OH}, a geometry optimization and frequency analysis was performed using Gaussian 09 \citep{gaussian09B1} at the B3LYP/6-311++G(d,p) level of theory and basis set.  The five lowest energy levels contribute $>$0.1\% correction to the partition function at 200~K, and have energies of 139, 188, 360, 396, and 584~cm$^{-1}$.  The total correction at 200~K is 2.49.

\section{\ce{CH3OH} Column Density}
\label{ch3oh_column}

\begin{figure*}
\centering
\includegraphics[width=\textwidth]{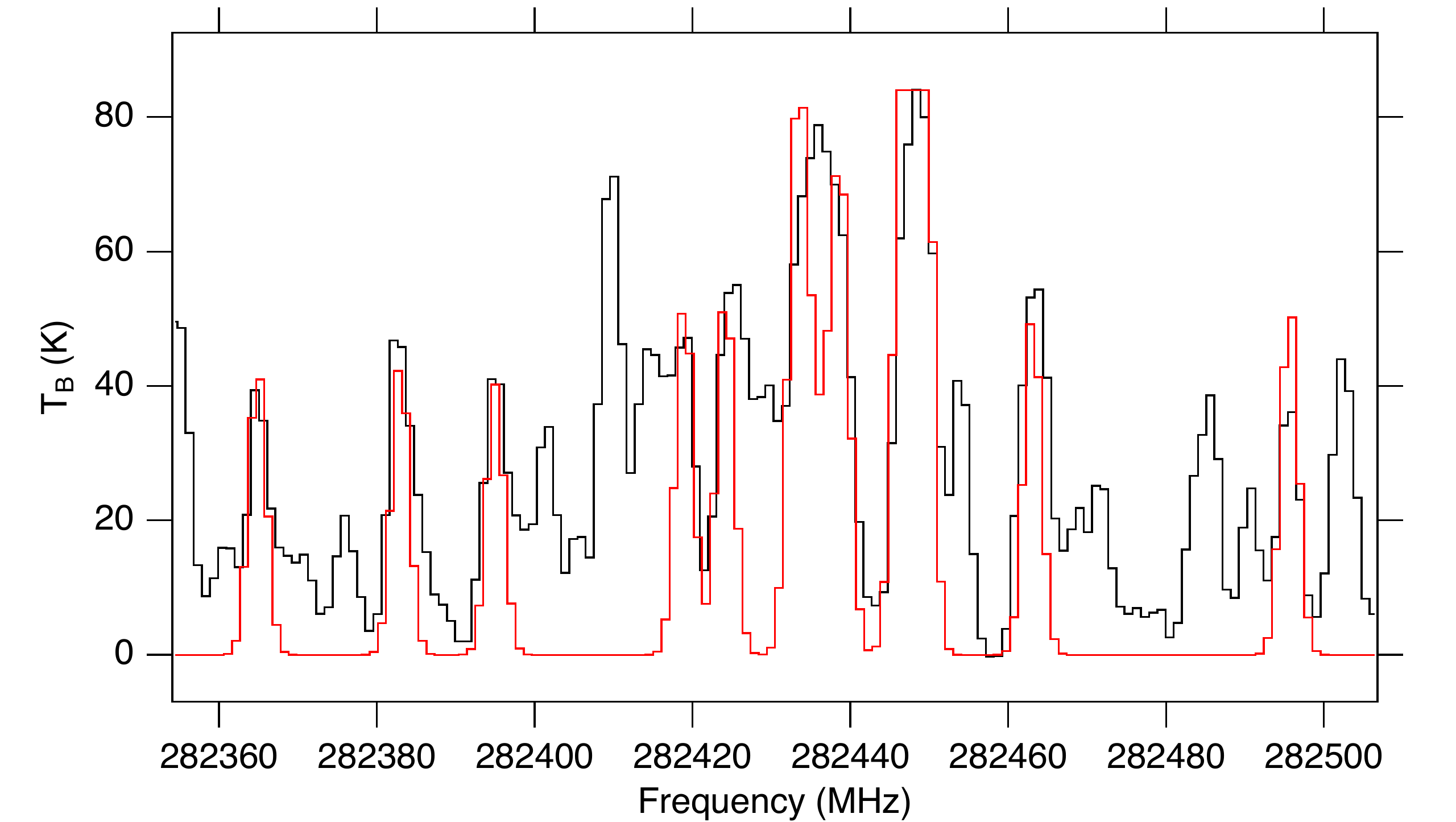}
\caption{Simulation of \ce{^{13}CH3OH}~v$_t$=1 in red over observations of MM1 in black with $N_T$~=~$1.95\times10^{18}$~cm$^{-2}$, T$_{ex}$~=~150~K, and $\Delta V$~=~2.8~\kms.}
\label{13ch3oh}
\end{figure*}

To compare with the column density of \ce{CH3OCH2OH}, a column density estimate of \ce{CH3OH} was required.  In the 0.2\arcsec~Band~7 data, nearly every \ce{CH3OH} transition is optically thick, as is nearly every transition of \ce{^{13}CH3OH}.  The lines of \ce{^{13}CH3OH}~v$_t$=1 are, however, largely optically thin (see Figure~\ref{13ch3oh}).  We assume that the v$_t$=0 and v$_t$=1 states of \ce{^{13}CH3OH} are described by a single excitation temperature, and calculate a total column density of \ce{^{13}CH3OH} using the $6_{3,4} - 5_{3,3}$ transition at 282383~MHz, which appears to be the least blended optically thin line.  A background temperature of $T_{bg}$~=~53~K was used, and an excitation temperature of $T_{ex}$~=~150~K, based on the background temperature and the brightness temperature of the optically thick \ce{^{13}CH3OH} lines.  The partition function included contributions from both v$_t$=0 and v$_t$=1.  A linewidth of 2.8~\kms~was fit to the lines.  The resulting column density was $N_T$~=~$1.95\times10^{18}$~cm$^{-2}$.  A \ce{CH3OH} column density of $N_T$~=~$1.4\times10^{20}$~cm$^{-2}$ was then inferred by scaling by a \ce{^{12}C}/\ce{^{13}C} ratio of 68 \citep{Milam:2005gt}.  Given the uncertainties involved and assumptions made, this value should be viewed as an estimate and as a lower limit.

\end{document}